\documentclass[reprint,amsmath,amssymb,aps,prl,nofootinbib]{revtex4-2}

\usepackage{graphicx}
\usepackage{dcolumn}
\usepackage{bm}
\usepackage{siunitx}
\usepackage{amsmath}
\usepackage{mathtools}
\usepackage{enumerate}
\usepackage{xcolor}

\begin{document}

\preprint{APS/123-QED}

\title{Scaling regimes for unsteady diffusion across particle-stabilized fluid interfaces}

\author{T.J.J.M. van Overveld}
\author{V. Garbin}
\email{v.garbin@tudelft.nl}
\affiliation{Department of Chemical Engineering, Delft University of Technology, Van der Maasweg 9, 2629 HZ Delft, The Netherlands
}

\date{\today}

\begin{abstract}
Colloidal particles at fluid interfaces can enhance the stability of drops and bubbles.
Yet, their effect on mass transfer in these multiphase systems remains ambiguous, with some experiments reporting strongly hindered diffusion, while others show nearly no effect, even at near-complete surface coverage.
To resolve this ambiguity, we solve the Fick-Jacobs equation for unsteady diffusion, allowing us to treat the particle-laden interface as a locally reduced cross-sectional area for mass transfer. Our numerical solutions reveal two limiting regimes, with the particle layer hindering diffusion only at short times. 
Guided by analytical solutions for a homogeneous layer with reduced diffusivity, we derive quantitative expressions for the transport regimes and associated transition times for diffusion across the particle layer.
This analysis yields a simple criterion for long-term hindrance that accurately distinguishes between conflicting experimental results, providing a unifying framework for mass transfer in particle-laden multiphase systems.
\end{abstract}

\maketitle

\begin{figure*}
    \centering
    \includegraphics[width=\linewidth]{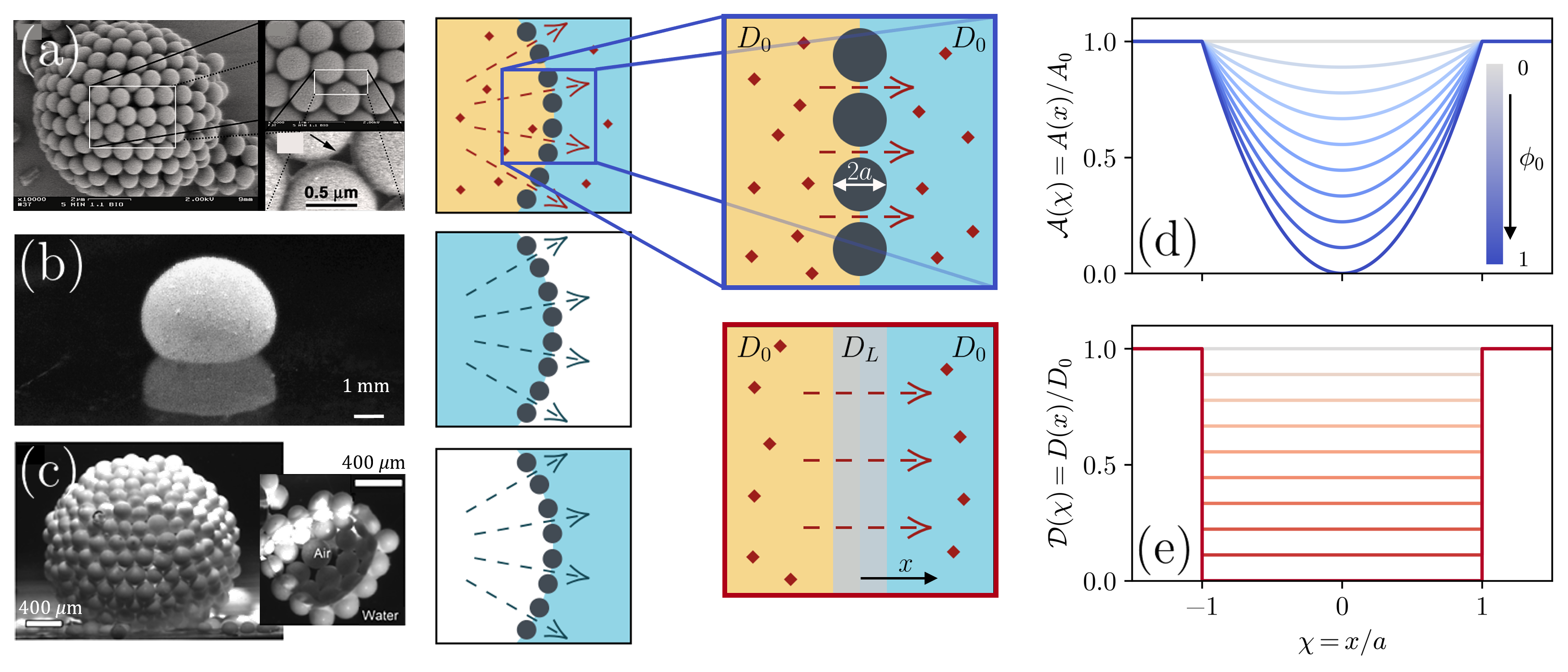}
    \caption{\textcolor{black}{Diffusion across particle-laden interfaces in multiphase systems.
    Diffusion between the inner and outer phases occurs through the interstitial pores between adsorbed particles, leading to processes such as 
    (a) solute exchange in colloidosomes (reprinted with permission from \citep{dinsmore2002colloidosomes}, copyright 2002 AAAS),
    (b) evaporation in liquid marbles (reprinted with permission from \citep{bormashenko2011liquid}, copyright 2011 Elsevier),
    and (c) gas diffusion in armored bubbles (reprinted with permission from \citep{subramaniam2006mechanics}, copyright 2006 ACS).
    The effect of a complex interface on diffusion can be modeled as
    (d) a layer of particles reducing the available cross-sectional area, depending on the surface coverage $\phi_0$, or (e) a homogeneous layer characterized by a locally reduced effective diffusivity.}}
    \label{fig:figure1}
\end{figure*}

\newpage
\textcolor{black}{Complex interfaces, stabilized by layers of molecules or solid particles, are ubiquitous in biological systems, manufactured materials and products, and in the environment \citep{sagis2011dynamic}. In particular,} colloidal particles adsorbed at liquid interfaces can stabilize a broad range of multiphase systems and lend long-term stability and functionality \citep{dinsmore2002colloidosomes,wu2016recent}, in applications ranging from foods and cosmetics to sustainable chemical conversion \citep{rousseau2013trends, marto2016green, crossley2010solid,rodriguez2020catalysis}.
\textcolor{black}{Their relatively high desorption energy compared to molecular surfactants imparts superior stability to emulsions and foams \citep{binks2002particles,marefati2020starch}.}
The semipermeability of particle-stabilized interfaces further allows for mass transfer between the different phases, leading to liquid evaporation from droplets \citep{bormashenko2011liquid}, gas dissolution from bubbles \citep{subramaniam2006mechanics}, or diffusive exchange of solutes in emulsions \citep{dan2012transport}, as shown in Fig.~\ref{fig:figure1}(a-c). 
\textcolor{black}{
Interfaces stabilized by colloidal particles also serve as useful model systems where the interfacial layer is more easily visualized compared to layers of molecules, allowing quantification of the microstructure and its evolution.}

\textcolor{black}{Despite their apparent simplicity}, the effect of interfacial particles on mass transfer remains ambiguous, with conflicting observations in the literature.  
At first glance, the reduced interfacial contact area should limit the net transport rate of solutes or solvents compared to uncoated droplets. Indeed, various experiments in both binary and ternary systems report that transport decreases as coverage increases \citep{sacca2008composition,dan2012transport,yong2016nanoparticle}. 
These findings contrast sharply with other studies showing no significant reduction in transport rate, even for densely covered interfaces \citep{thompson2010covalently,schroder2019can,prakash2025evaporation}. 
A notable delay in solute release from colloidosome microcapsules \citep{dinsmore2002colloidosomes} is only observed after processing the layer to \textcolor{black}{increase the surface coverage beyond the close-packing limit of spheres, often reducing} pore sizes beyond the resolution of electron microscopy \citep{yow2009release, thompson2010covalently}.
Our group previously measured the spatio-temporal evolution of solute concentration around Pickering droplets, confirming that the concentration field is only affected near the particle layer, while the far-field evolution remains similar to that of a bare interface \cite{liu2024diffusion}. Similar trends have been observed in porous media, where reduced porosity or effective diffusivity does not necessarily lead to a proportional decrease in mass transfer \citep{suzuki1968mechanism,shahraeeni2012coupling}.


The effect of particle layers on diffusion is typically interpreted using the seminal framework of \citet{berg1993random}, which describes steady-state diffusion across porous membranes via an effective interfacial resistance. 
This model predicts that diffusion is barely hindered for experimentally relevant pore sizes, and significant resistance requires extremely narrow pores.
However, the framework assumes constant concentration profiles in equilibrium with infinite reservoirs, which is rarely met in practice. Most experiments involve transient diffusion from finite droplets with dynamically evolving concentrations \citep{thompson2010covalently,yow2009release,sacca2008composition,dan2012transport,yong2016nanoparticle,schroder2019can,prakash2025evaporation}, which consequently alter the interfacial fluxes \citep{liu2024diffusion}. In such systems, especially with densely covered interfaces, diffusion can be so strongly impeded that steady state is never reached within the experimental time frame. A transient model is therefore essential for capturing non-equilibrium behavior and quantifying the associated transport regimes.

In this Letter, we reconcile the apparent contradictions in the effect of particle-laden interfaces on mass transfer, using an unsteady diffusion model that accounts for the particles through a reduction in available area. The resulting diffusive transport closely resembles that across a homogeneous layer with reduced effective diffusivity.
Through this analogy, we demonstrate how diffusion can appear unaffected by a particle layer or barrier, even when local diffusivity or interfacial area is significantly reduced.
\textcolor{black}{Even though our simplified model does not account for detailed physicochemical processes, it captures a dominant mechanism that} further unifies transport phenomena across diverse systems, including multicomponent droplet evaporation and vapor dissolution, as shown in Fig.~\ref{fig:figure1}.

A complex interface may consist of a layer of particles of radius $a$, which reduces the effective cross-sectional area $A(x)$ available for diffusion at position $x$ compared to the reference area $A_0$ outside the layer, as shown in Fig.~\ref{fig:figure1}(d). An interfacial layer (of thickness $2a$) may also have a spatially varying diffusivity $D(x)$, deviating from the bulk reference diffusivity $D_0$, as in Fig.~\ref{fig:figure1}(e).
Both effects on diffusive transport are combined in the Fick-Jacobs equation,
\begin{equation}\label{eq:FickJacobs_dimless}
    \frac{\partial c(\chi,\tau)}{\partial \tau} = \frac{1}{\mathcal{A}(\chi)}\frac{\partial}{\partial \chi}\left[\mathcal{A}(\chi)\mathcal{D}(\chi)\frac{\partial c(\chi,\tau)}{\partial \chi}\right], \\
\end{equation}
\textcolor{black}{where $c(\chi,\tau)$ is the normalized solute concentration averaged over the cross-sectional plane\footnote{\textcolor{black}{Note that this formalism is equivalent to the one used in \citep{zwanzig1992diffusion,reguera2001kinetic,kalinay2006corrections}, based on the total amount of solute in each cross-sectional plane.}} at dimensionless position $\chi=x/a$ and dimensionless time $\tau=D_0t/a^2$, according to the characteristic timescale for diffusion over a layer, $a^2/D_0$. 
The right-hand side accounts for variations in the diffusive flux due to changes in the available area $\mathcal{A}(\chi)=A(x)/A_0$, diffusion coefficient $\mathcal{D}(\chi)=D(x)/D_0$, and concentration gradient (i.e., Fickian diffusion) \citep{jacobs1935diffusion}.}
The one-dimensional \textcolor{black}{formulation} in Eq.~\eqref{eq:FickJacobs_dimless} applies to the three-dimensional concentration field around a particle-laden interface (Fig.~\ref{fig:figure1}(d)), assuming rapid equilibration parallel to the interface. 
\textcolor{black}{This approximation has previously been validated against full three-dimensional numerical simulations \citep{liu2024diffusion}, and is further supported by two-dimensional simulations with equivalent area reduction (see the Supplemental Material).}
 
For an interface stabilized by a monolayer of spherical particles, the cross-sectional area varies as 
\begin{equation}\label{eq:g(x)}
    \mathcal{A}(\chi) = 
    \begin{dcases}
        1-\phi_0\left(1-\chi^2\right) & \mathrm{for} -1\leq \chi\leq 1, \\
        1 & \mathrm{otherwise},\\
    \end{dcases}
\end{equation}
where $\phi_0$ is the packing fraction at the particle layer midplane, \textcolor{black}{up to the circle packing limit ($\phi\approx0.91$) \citep{liu2024diffusion}. The quantity $\phi_0$ can be interpreted more generally as the surface coverage for any regular particle geometry at the interface, including non-spherical particles, effectively allowing values exceeding the dense circle packing limit, with the functional form of $\mathcal{A}(\chi)$ determined by the particle shape. 
The Fick-Jacobs framework requires a correction to the effective diffusivity when $\mathcal{A}(\chi)$ varies rapidly. We use a correction based on the most accurate expression attainable using first derivatives of the cross-sectional area \citep{zwanzig1992diffusion,reguera2001kinetic,kalinay2006corrections}, yielding
\begin{equation}\label{eq:DiffusivityCorrection}
    \mathcal{D}(\chi)=
    \begin{dcases}
        \sqrt{\dfrac{1-\phi_0(1-\chi^2)}{1-\phi_0(1-2\chi^2)}} & \mathrm{for} -1\leq \chi\leq 1, \\
        1 & \mathrm{otherwise},
    \end{dcases}
\end{equation}
as detailed in the Supplemental Material.}
The effect of the particle-laden interface on the evolution of the concentration profile is obtained by numerically solving Eq.~\eqref{eq:FickJacobs_dimless} for various surface coverage $\phi_0$, using a finite-difference scheme. We initialize the concentration field with a step from a solute-rich to a solute-free phase across the interface. 

\begin{figure*}
    \centering
    \includegraphics[width=\linewidth]{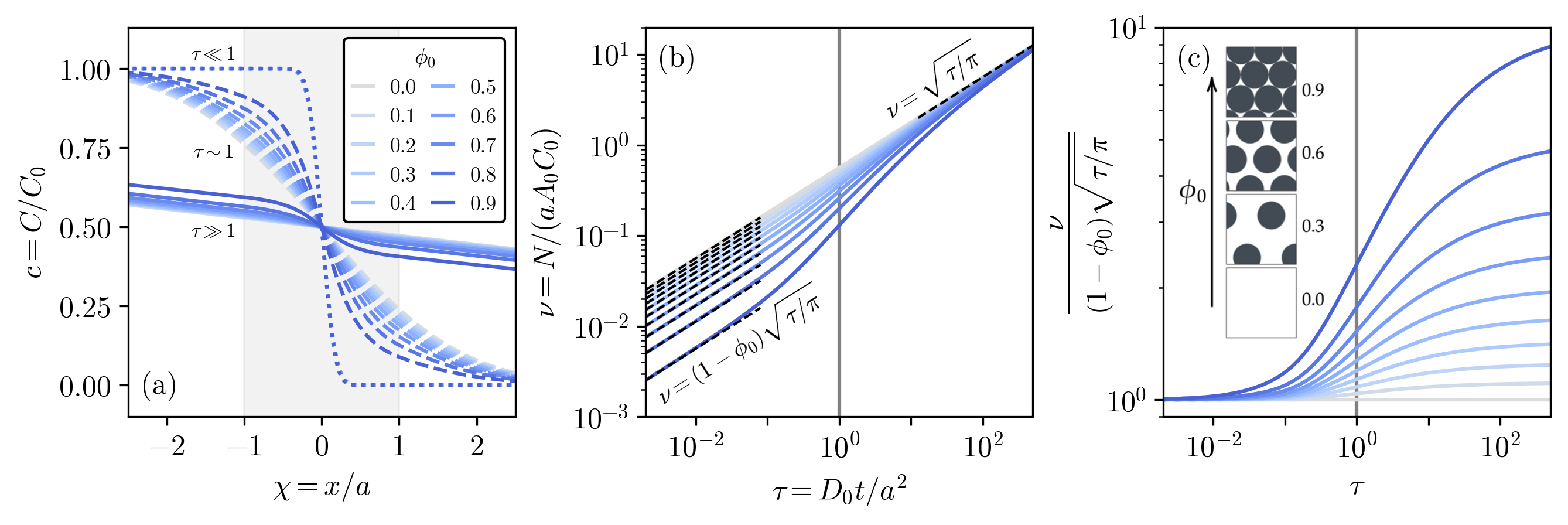}
    \caption{Numerical results for unsteady diffusion across a particle-laden interface. (a) Concentration profiles evolve identically at short times ($\tau\ll 1$, \textcolor{black}{dotted curves}), but steeper gradients develop within the layer for higher surface coverage $\phi_0$ \textcolor{black}{at later times (dashed and solid curves)}.
    (b) The diffused solute mass normalized by the layer volume, $\nu$, is initially reduced by a factor of $1-\phi_0$ compared to the bare interface case. At long times, all curves converge to the bare interface solution $\sqrt{\tau/\pi}$.
    (c) The regime transitions are not well captured by simply rescaling the diffused mass $\nu$ by the short-time asymptotes. 
    }
    \label{fig:figure2}
\end{figure*}

The solute concentration profiles around the particle layer show no dependence on $\phi_0$ at very short times (${\tau\ll1}$), while at intermediate (${\tau\sim1}$) and long times (${\tau\gg1}$), the surface coverage significantly affects the gradients within the layer, shown in Fig.~\ref{fig:figure2}(a).
To investigate these transport regimes, we compute the total mass of solute that has crossed the interface, $N$, from the concentration gradient at the interface. The normalized solute mass, $\nu=N/(aA_0C_0)$, is plotted in Fig.~\ref{fig:figure2}(b), highlighting two limiting regimes.

At long times, all curves converge to the bare-interface solution $\nu=\sqrt{\tau/\pi}$, surprisingly indicating that the particle layer no longer affects the overall transport.
In contrast, at short times, the surface coverage reduces transport, resulting in lower values of $\nu$, while the initial slopes remain unaffected across cases.  
The dominant effect of the particle layer at short times is due to diffusion being confined to a narrow region near the interface, with self-similar concentration profiles within the individual pores between particles. 
We find that the value of $\nu$ in this regime is directly proportional to the available interfacial area, which is reduced by a factor $1-\phi_0$ compared to a bare interface. Rescaling the vertical axis by this short-time asymptote in Fig.~\ref{fig:figure2}(c), however, reveals a collapse that only holds for very short times ($\tau\lesssim10^{-2}$). 
\textcolor{black}{The transition between regimes is clearly not captured by the dimensionless time $\tau$, despite it being the \emph{a priori} natural choice for scaling time.
Moreover, cases with higher $\phi_0$ values deviate from the short-time asymptote earlier, yet take longer to reach the long-term bare-interface asymptote. 
Because $\phi_0$ is dimensionless, dimensional analysis alone does not yield a quantitative scaling law.} 

To understand quantitatively the nontrivial effect of surface coverage on the evolution of solute diffusion, we analyze the limiting behavior of a closely related system for which analytical solutions to Eq.~\eqref{eq:FickJacobs_dimless} are known.
In this system, the particles are modeled as a homogeneous layer with a constant diffusivity $D_L < D_0$ (cf. Fig.~\ref{fig:figure1}(e)), effectively accounting for the hindrance to mass transfer via reduced diffusivity instead of reduced available area.
\begin{figure*}
    \centering
    \includegraphics[width=\textwidth]{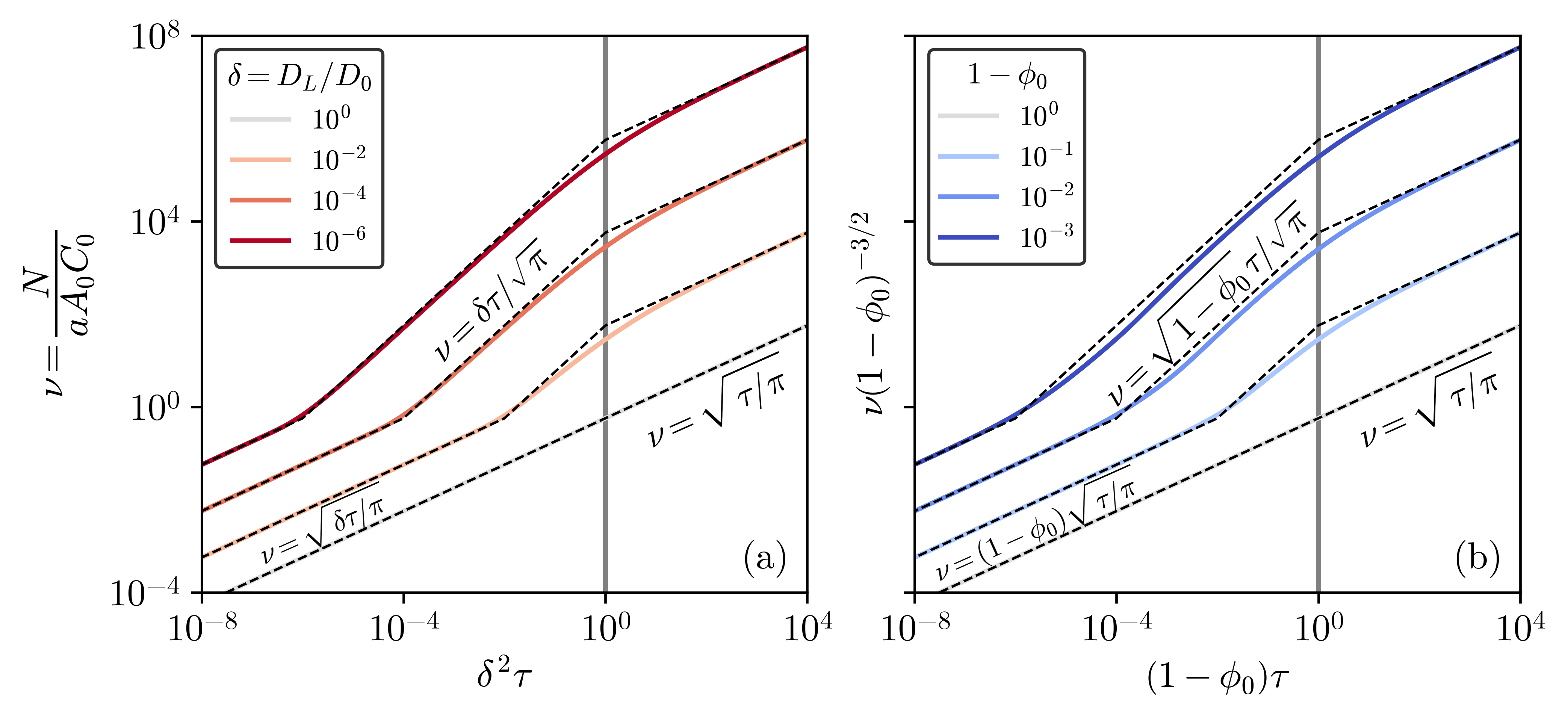}
    \caption{Unsteady diffusion across a homogeneous layer with reduced diffusivity as an effective model for a particle-laden interface.
    (a) Analytical solutions for a layer with reduced diffusivity [Eq.~\eqref{eq:NaAC}]. (b) Numerical solutions of the Fick-Jacobs equation [Eq.~\eqref{eq:FickJacobs_dimless}] for a particle-laden interface.
    In both plots, time is normalized by the characteristic timescale that separates the intermediate and long-time regimes. The vertical axis in (b) is scaled by $(1-\phi_0)^{3/2}$ to account for the reduced fluid volume within the layer, aligning the curves with the analytical results.
    Black dashed lines represent first-order approximations for the transport regimes based on Eqs.~\eqref{eq:NaAC_approx}~and~\eqref{eq:nu_phi0_approx}.}
    \label{fig:figure3}
\end{figure*}
From the analytical solutions for $c(\chi,\tau)$ \citep{CarslawJaeger1959}, we calculate the normalized mass of solute that has crossed the interface, resulting in
\begin{equation}\label{eq:NaAC}
\begin{split}
    \nu(\tau,\delta) &= \sqrt{\frac{\delta\tau}{\pi}}\left(1+2\sum\limits_{n=1}^\infty\alpha^n e^{-n^2/(\delta\tau)}\right) \\
    &\qquad\qquad - 2\sum\limits_{n=1}^\infty n\alpha^n \mathrm{erfc}\left(\frac{n}{\sqrt{\delta\tau}}\right), \\
\end{split}
\end{equation}
with parameters $\delta=D_L/D_0$ and $\alpha=(1-\sqrt{\delta})/(1+\sqrt{\delta})$, shown as solid lines for different values of $\delta$ in Fig.~\ref{fig:figure3}(a).
Asymptotic analysis of Eq.~\eqref{eq:NaAC} reveals limiting regimes at short and long times, connected by an intermediate regime
\textcolor{black}{\begin{equation}\label{eq:NaAC_approx}
\nu \approx \dfrac{\delta\tau}{\sqrt{\pi}} 
\end{equation}
for $1/\delta < \tau <1/\delta^2$. These scaling regimes are shown as dashed lines in Fig.~\ref{fig:figure3}(a), where the horizontal axis is scaled to highlight the transition between the intermediate and the long-time regime at $\delta^2\tau=1$. The linear dependence of $\nu$ on $\tau$ in the intermediate regime is particularly evident for small values of $\delta$.}

The analytical solutions for the homogeneous layer with reduced diffusivity now guide the scaling analysis for the particle-laden interface. 
While the earlier rescaling in Fig.~\ref{fig:figure2}(c) failed to capture the regime transitions, the new insight into the intermediate regime enables a more effective rescaling of the numerical solutions for different values of $\phi_0$, shown as solid lines in Fig.~\ref{fig:figure3}(b). This rescaling reveals the same three regimes observed in the analytical solutions of Fig.~\ref{fig:figure3}(a).
The time $\tau$ is rescaled by $1-\phi_0$ in Fig.~\ref{fig:figure3}(b), which aligns the transition from the intermediate to the long-time regime at $\tau=1/(1-\phi_0)$. 
Simultaneously, the total diffused mass $\nu$ is rescaled by $(1-\phi_0)^{3/2}$, consistent with the reduced fluid volume due to the particle presence.
The rescaling of both axes explains why the earlier scaling using only $1-\phi_0$ in Fig.~\ref{fig:figure2}(c) was unsuccessful.
Moreover, the vertical rescaling reveals an additional timescale in the Fick-Jacobs framework associated with diffusion through the layer itself, aligning the transition between short and intermediate time regimes, as shown in Fig.~S1 in the Supplemental Material.

The strong similarity of the numerical solutions for a particle-laden interface with the corresponding analytical solutions for a homogeneous layer [Figs.~\ref{fig:figure3}(a-b)] suggests the following approximate scaling for diffusion across the particle-laden interface \textcolor{black}{in the intermediate regime:
\begin{equation}\label{eq:nu_phi0_approx}
    \nu(\tau,\phi_0) \approx \tau \sqrt{\frac{1-\phi_0}{\pi}}\,,
\end{equation}
valid for $1-\phi_0 < \tau < 1/(1-\phi_0)$.
The scaling regimes are shown as dashed lines in Fig.~\ref{fig:figure3}(b), centered around the transition to the long-time regime at $(1-\phi_0)\tau=1$.}
\textcolor{black}{The 1D model is valid for homogeneous layers; a heterogeneous pore size distribution, leading to spanwise asymmetries in the concentration field around the layer, would cause a more gradual transition between regimes. The effect should be especially pronounced at short times, when the layer properties dominate the transport.
}
The comparison between the two frameworks also reveals that there is no single universal conversion between effective diffusivity and interfacial coverage. Nonetheless, a regime-specific conversion by 
yields
\begin{equation}\label{eq:conversion}
    D_L \approx \left\{
    \begin{aligned}
        D_0&(1-\phi_0)^2 \quad &\mathrm{for\ \ }& &\tau < 1-\phi_0, \\
        D_0&\sqrt{1-\phi_0} \quad &\mathrm{for\ \ }& 1-\phi_0 < &\tau< \dfrac{1}{1-\phi_0}, \\
        D_0& \quad &\mathrm{for\ \ }& &\tau>\dfrac{1}{1-\phi_0}.
    \end{aligned}
    \right.
\end{equation}

With the quantitative expressions for transport regimes across a particle-laden interface, we can now reconcile the conflicting observations in the literature, ranging from long-term hindrance to virtually unaffected transport in Pickering emulsions and evaporating droplets \citep{sacca2008composition,dan2012transport,yong2016nanoparticle,thompson2010covalently,schroder2019can,prakash2025evaporation,yow2009release,liu2024diffusion}.
The key parameter is the characteristic timescale 
\begin{equation}\label{eq:tl}
    t_L=\frac{a^2}{D_0\left(1-\phi_0\right)},
\end{equation}
which marks the transition from the intermediate regime, where solute diffusion is hindered by the interface, to the long-time regime, where the effect of the interface becomes negligible. 
For a system of characteristic size $R$ (e.g., droplet radius), diffusion across the system occurs on the timescale $t_R=R^2/D_0$. If $t_L\ll t_R$, the influence of the interface vanishes well before diffusion on the system-scale becomes relevant. Conversely, if $t_L\gtrsim t_R$, solute transport remains hindered well beyond $t_R$. This comparison yields a simple criterion for long-term hindrance of diffusive transport 
\begin{equation}\label{eq:criterion}
    \phi_0 > 1-\frac{a^2}{R^2}, 
\end{equation}
which can be equivalently expressed in terms of the effective diffusivity as
\begin{equation}
    \frac{D_L}{D_0}<\frac{a}{R}.
\end{equation}

\begin{figure}
    \centering
    \includegraphics[width=\linewidth]{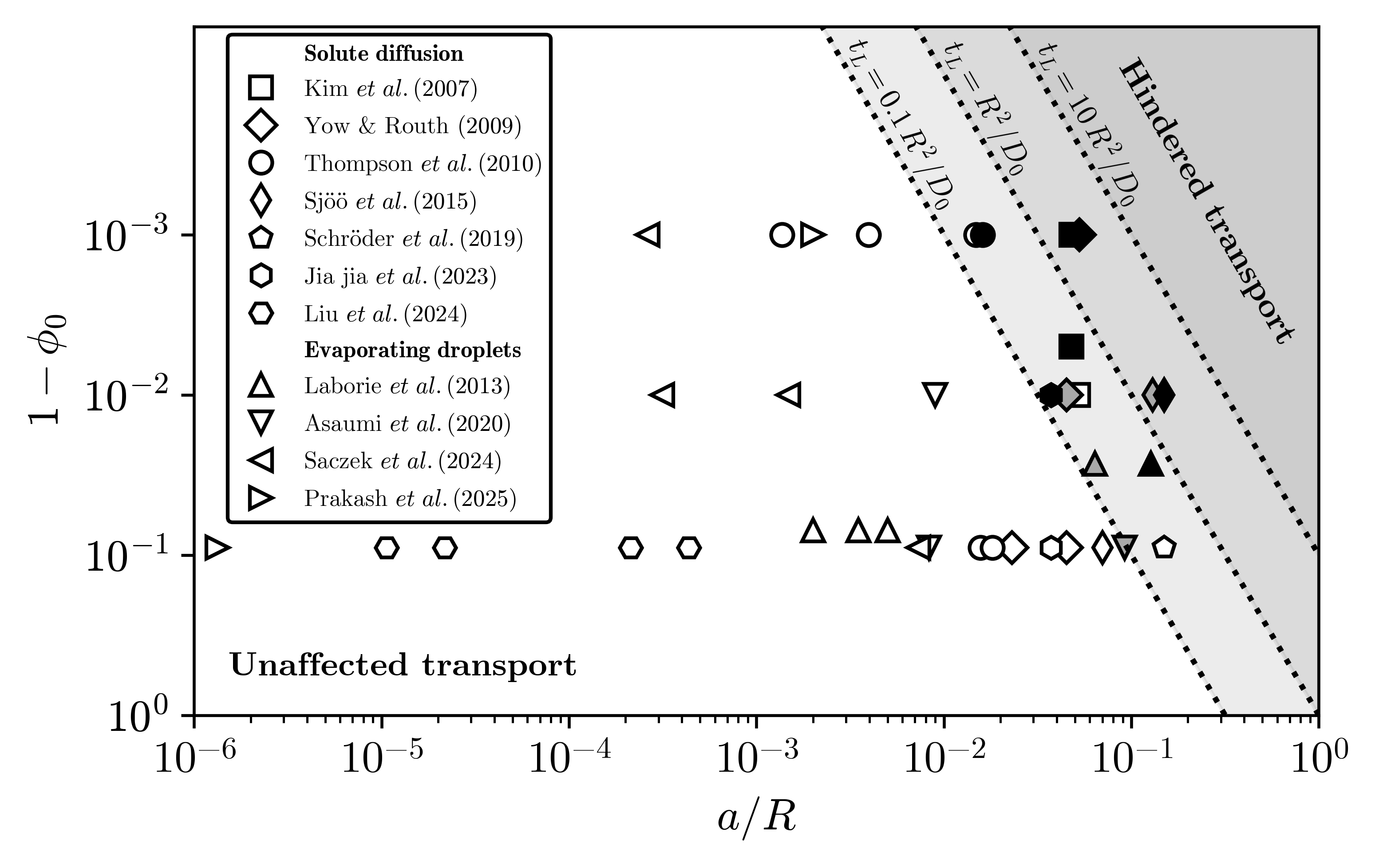}
    \caption{Data from the literature compared against the criterion for long-term hindrance to diffusion.
    Experimental data includes evaporating droplets (triangles) \citep{laborie2013coatings,asaumi2020effect,saczek2024impact,prakash2025evaporation} and solute diffusion in emulsions (other symbols) \citep{kim2007colloidal, yow2009release, thompson2010covalently, sjoo2015barrier,schroder2019can,jia2023efficient,liu2024diffusion}, colored by the observed effect of the particle-laden interface: significant hindrance (black), minor effect (gray), or no effect (white).
    Dashed lines represent the criterion from Eq.~\eqref{eq:criterion} and offsets by factors of ten, reflecting the typical sensitivity of our analysis. Together, these lines mark the transition from unaffected to strongly hindered diffusion.} 
    \label{fig:figure4}
\end{figure}

The criterion for effective hindrance of diffusive transport is shown in the $(a/R, 1-\phi_0)$ space in Fig.~\ref{fig:figure4}, alongside experimental data from various systems, where either surface coverage or effective layer diffusivity is reported. 
The present formulation [Eq.~\eqref{eq:criterion}] enables direct comparison across systems without requiring the often unavailable value of $D_0$, while $R$ is typically reported or readily estimated from experimental data.
For systems where direct layer visualizations suggest dense circle packing, we typically assume $\phi_0\approx 0.91$. For systems with processing steps that further reduce pore area, accurately estimating the surface coverage from layer visualizations is challenging. 
\textcolor{black}{
Comparing typical pore sizes at various values of $\phi_0$ to representative micrographs from the literature \citep{dinsmore2002colloidosomes,yow2009release,thompson2010covalently} suggests that the available free area $1-\phi_0$ decreases by roughly one order of magnitude per processing step. This estimation keeps $\phi_0$ values consistent within experimental series from the same study. 
We cannot reliably assign values of $1-\phi_0$ below $10^{-3}$, as this typically corresponds to pore sizes below the imaging resolution. While the actual available interfacial area may be smaller, it is not readily quantifiable. We therefore project the cases with $1-\phi_0\lesssim10^{-3}$ onto $10^{-3}$. } 
For evaporating droplets, the reported values of $\phi_0$ typically refer to initial states and are assumed to remain approximately constant during early-time evaporation, when the structure of the interfacial particle layer, and its influence on evaporation, is assessed.
\textcolor{black}{Detailed calculations for each experiment are shown in the Supplemental Material, where we additionally confirm that the effective pore radius exceeds the hydrodynamic radius of the diffusing solute, ruling out effects due to size exclusion.}
Systems showing hindered transport due to the interfacial layer (filled symbols) predominantly lie in the region where $t_L>R^2/D_0$ (shaded area), while cases where the layer has no effect fall well outside this region.
\textcolor{black}{Note that the threshold remains inconclusive for some cases, particularly the gray and white symbols within the shaded areas. This may reflect the uncertainties in estimating layer thickness and surface coverage from specific experiments, and could also suggest that additional physical effects are relevant in those cases. Nonetheless, the}
parameter space overview highlights that effective hindered transport requires interfaces that are both sufficiently thick (typically $a/R\gtrsim0.01$) and very densely covered (with most data points for $\phi_0 \gtrsim 0.99$).
For particle monolayers, this demands exceptionally high surface coverage, well above the dense circle packing limit $\phi\gg0.91$. In the experiments reported in Fig.~\ref{fig:figure4}, this is achieved, for instance, by thermal sintering \citep{yow2009release} or polymer deposition \citep{thompson2010covalently}.
For armored bubbles, the literature typically reports only the late stages of the system, after dissolution arrest \citep{subramaniam2006mechanics, abkarian2007dissolution} or buckling \citep{taccoen2016probing,poulichet2015cooling}, providing insufficient data to include them in our analysis in Fig.~\ref{fig:figure4}. 

The results from our transient approach are consistent with Berg's steady-state solutions: both frameworks predict that diffusive transport is significantly hindered for very high surface coverage \citep{berg1993random}. 
Yet, a transient approach is especially important for well-covered interfaces, where the timescale $t_L$, marking the onset of bulk-dominated diffusion, can exceed the layer diffusion timescale $a^2/D_L$ by multiple orders of magnitude. As a result, transients may persist throughout all experimentally relevant time scales, rendering steady-state assumptions invalid.
This time-resolved perspective \textcolor{black}{qualitatively} distinguishes between experiments with and without long-term transport hindrance. \textcolor{black}{While such a simplified model cannot capture the full transient dynamics of each system, \textcolor{black}{the results highlight the major effect of diffusion to explain} the contrasting observations in the literature.
As a future direction, more accurate models including specific physico-chemical properties of emulsions, evaporating drops, or dissolving bubbles will enable more quantitative descriptions of mass transfer across complex interfaces. }
More broadly, the model is generally applicable to other complex interfaces ubiquitous in nature and technology.



\begin{acknowledgments}
We thank B.P. Binks, A.G. Mar\'{i}n, and A.T. Oratis for insightful discussions.
This work is funded by the Dutch Research Council (NWO) under grant number 19929.
\end{acknowledgments}

The data that support the findings of this article are openly available \citep{vanoverveld2025dataset}.


\begin{thebibliography}{37}%
\makeatletter
\providecommand \@ifxundefined [1]{%
 \@ifx{#1\undefined}
}%
\providecommand \@ifnum [1]{%
 \ifnum #1\expandafter \@firstoftwo
 \else \expandafter \@secondoftwo
 \fi
}%
\providecommand \@ifx [1]{%
 \ifx #1\expandafter \@firstoftwo
 \else \expandafter \@secondoftwo
 \fi
}%
\providecommand \natexlab [1]{#1}%
\providecommand \enquote  [1]{``#1''}%
\providecommand \bibnamefont  [1]{#1}%
\providecommand \bibfnamefont [1]{#1}%
\providecommand \citenamefont [1]{#1}%
\providecommand \href@noop [0]{\@secondoftwo}%
\providecommand \href [0]{\begingroup \@sanitize@url \@href}%
\providecommand \@href[1]{\@@startlink{#1}\@@href}%
\providecommand \@@href[1]{\endgroup#1\@@endlink}%
\providecommand \@sanitize@url [0]{\catcode `\\12\catcode `\$12\catcode `\&12\catcode `\#12\catcode `\^12\catcode `\_12\catcode `\%12\relax}%
\providecommand \@@startlink[1]{}%
\providecommand \@@endlink[0]{}%
\providecommand \url  [0]{\begingroup\@sanitize@url \@url }%
\providecommand \@url [1]{\endgroup\@href {#1}{\urlprefix }}%
\providecommand \urlprefix  [0]{URL }%
\providecommand \Eprint [0]{\href }%
\providecommand \doibase [0]{https://doi.org/}%
\providecommand \selectlanguage [0]{\@gobble}%
\providecommand \bibinfo  [0]{\@secondoftwo}%
\providecommand \bibfield  [0]{\@secondoftwo}%
\providecommand \translation [1]{[#1]}%
\providecommand \BibitemOpen [0]{}%
\providecommand \bibitemStop [0]{}%
\providecommand \bibitemNoStop [0]{.\EOS\space}%
\providecommand \EOS [0]{\spacefactor3000\relax}%
\providecommand \BibitemShut  [1]{\csname bibitem#1\endcsname}%
\let\auto@bib@innerbib\@empty
\bibitem [{\citenamefont {Sagis}(2011)}]{sagis2011dynamic}%
  \BibitemOpen
  \bibfield  {author} {\bibinfo {author} {\bibfnamefont {L.~M.}\ \bibnamefont {Sagis}},\ }\bibfield  {title} {\bibinfo {title} {Dynamic properties of interfaces in soft matter: Experiments and theory},\ }\href@noop {} {\bibfield  {journal} {\bibinfo  {journal} {Reviews of Modern Physics}\ }\textbf {\bibinfo {volume} {83}},\ \bibinfo {pages} {1367} (\bibinfo {year} {2011})}\BibitemShut {NoStop}%
\bibitem [{\citenamefont {Dinsmore}\ \emph {et~al.}(2002)\citenamefont {Dinsmore}, \citenamefont {Hsu}, \citenamefont {Nikolaides}, \citenamefont {Marquez}, \citenamefont {Bausch},\ and\ \citenamefont {Weitz}}]{dinsmore2002colloidosomes}%
  \BibitemOpen
  \bibfield  {author} {\bibinfo {author} {\bibfnamefont {A.}~\bibnamefont {Dinsmore}}, \bibinfo {author} {\bibfnamefont {M.~F.}\ \bibnamefont {Hsu}}, \bibinfo {author} {\bibfnamefont {M.}~\bibnamefont {Nikolaides}}, \bibinfo {author} {\bibfnamefont {M.}~\bibnamefont {Marquez}}, \bibinfo {author} {\bibfnamefont {A.}~\bibnamefont {Bausch}},\ and\ \bibinfo {author} {\bibfnamefont {D.}~\bibnamefont {Weitz}},\ }\bibfield  {title} {\bibinfo {title} {Colloidosomes: selectively permeable capsules composed of colloidal particles},\ }\href@noop {} {\bibfield  {journal} {\bibinfo  {journal} {Science}\ }\textbf {\bibinfo {volume} {298}},\ \bibinfo {pages} {1006} (\bibinfo {year} {2002})}\BibitemShut {NoStop}%
\bibitem [{\citenamefont {Wu}\ and\ \citenamefont {Ma}(2016)}]{wu2016recent}%
  \BibitemOpen
  \bibfield  {author} {\bibinfo {author} {\bibfnamefont {J.}~\bibnamefont {Wu}}\ and\ \bibinfo {author} {\bibfnamefont {G.-H.}\ \bibnamefont {Ma}},\ }\bibfield  {title} {\bibinfo {title} {Recent studies of pickering emulsions: particles make the difference},\ }\href@noop {} {\bibfield  {journal} {\bibinfo  {journal} {Small}\ }\textbf {\bibinfo {volume} {12}},\ \bibinfo {pages} {4633} (\bibinfo {year} {2016})}\BibitemShut {NoStop}%
\bibitem [{\citenamefont {Rousseau}(2013)}]{rousseau2013trends}%
  \BibitemOpen
  \bibfield  {author} {\bibinfo {author} {\bibfnamefont {D.}~\bibnamefont {Rousseau}},\ }\bibfield  {title} {\bibinfo {title} {Trends in structuring edible emulsions with pickering fat crystals},\ }\href@noop {} {\bibfield  {journal} {\bibinfo  {journal} {Current Opinion in Colloid \& Interface Science}\ }\textbf {\bibinfo {volume} {18}},\ \bibinfo {pages} {283} (\bibinfo {year} {2013})}\BibitemShut {NoStop}%
\bibitem [{\citenamefont {Marto}\ \emph {et~al.}(2016)\citenamefont {Marto}, \citenamefont {Gouveia}, \citenamefont {Chiari}, \citenamefont {Paiva}, \citenamefont {Isaac}, \citenamefont {Pinto}, \citenamefont {Sim{\~o}es}, \citenamefont {Almeida},\ and\ \citenamefont {Ribeiro}}]{marto2016green}%
  \BibitemOpen
  \bibfield  {author} {\bibinfo {author} {\bibfnamefont {J.}~\bibnamefont {Marto}}, \bibinfo {author} {\bibfnamefont {L.}~\bibnamefont {Gouveia}}, \bibinfo {author} {\bibfnamefont {B.}~\bibnamefont {Chiari}}, \bibinfo {author} {\bibfnamefont {A.}~\bibnamefont {Paiva}}, \bibinfo {author} {\bibfnamefont {V.}~\bibnamefont {Isaac}}, \bibinfo {author} {\bibfnamefont {P.}~\bibnamefont {Pinto}}, \bibinfo {author} {\bibfnamefont {P.}~\bibnamefont {Sim{\~o}es}}, \bibinfo {author} {\bibfnamefont {A.}~\bibnamefont {Almeida}},\ and\ \bibinfo {author} {\bibfnamefont {H.}~\bibnamefont {Ribeiro}},\ }\bibfield  {title} {\bibinfo {title} {The green generation of sunscreens: Using coffee industrial sub-products},\ }\href@noop {} {\bibfield  {journal} {\bibinfo  {journal} {Industrial Crops and Products}\ }\textbf {\bibinfo {volume} {80}},\ \bibinfo {pages} {93} (\bibinfo {year} {2016})}\BibitemShut {NoStop}%
\bibitem [{\citenamefont {Crossley}\ \emph {et~al.}(2010)\citenamefont {Crossley}, \citenamefont {Faria}, \citenamefont {Shen},\ and\ \citenamefont {Resasco}}]{crossley2010solid}%
  \BibitemOpen
  \bibfield  {author} {\bibinfo {author} {\bibfnamefont {S.}~\bibnamefont {Crossley}}, \bibinfo {author} {\bibfnamefont {J.}~\bibnamefont {Faria}}, \bibinfo {author} {\bibfnamefont {M.}~\bibnamefont {Shen}},\ and\ \bibinfo {author} {\bibfnamefont {D.~E.}\ \bibnamefont {Resasco}},\ }\bibfield  {title} {\bibinfo {title} {Solid nanoparticles that catalyze biofuel upgrade reactions at the water/oil interface},\ }\href@noop {} {\bibfield  {journal} {\bibinfo  {journal} {Science}\ }\textbf {\bibinfo {volume} {327}},\ \bibinfo {pages} {68} (\bibinfo {year} {2010})}\BibitemShut {NoStop}%
\bibitem [{\citenamefont {Rodriguez}\ and\ \citenamefont {Binks}(2020)}]{rodriguez2020catalysis}%
  \BibitemOpen
  \bibfield  {author} {\bibinfo {author} {\bibfnamefont {A.~M.~B.}\ \bibnamefont {Rodriguez}}\ and\ \bibinfo {author} {\bibfnamefont {B.~P.}\ \bibnamefont {Binks}},\ }\bibfield  {title} {\bibinfo {title} {Catalysis in pickering emulsions},\ }\href@noop {} {\bibfield  {journal} {\bibinfo  {journal} {Soft Matter}\ }\textbf {\bibinfo {volume} {16}},\ \bibinfo {pages} {10221} (\bibinfo {year} {2020})}\BibitemShut {NoStop}%
\bibitem [{\citenamefont {Binks}(2002)}]{binks2002particles}%
  \BibitemOpen
  \bibfield  {author} {\bibinfo {author} {\bibfnamefont {B.~P.}\ \bibnamefont {Binks}},\ }\bibfield  {title} {\bibinfo {title} {Particles as surfactants—similarities and differences},\ }\href@noop {} {\bibfield  {journal} {\bibinfo  {journal} {Current opinion in colloid \& interface science}\ }\textbf {\bibinfo {volume} {7}},\ \bibinfo {pages} {21} (\bibinfo {year} {2002})}\BibitemShut {NoStop}%
\bibitem [{\citenamefont {Marefati}\ and\ \citenamefont {Rayner}(2020)}]{marefati2020starch}%
  \BibitemOpen
  \bibfield  {author} {\bibinfo {author} {\bibfnamefont {A.}~\bibnamefont {Marefati}}\ and\ \bibinfo {author} {\bibfnamefont {M.}~\bibnamefont {Rayner}},\ }\bibfield  {title} {\bibinfo {title} {Starch granule stabilized pickering emulsions: an 8-year stability study},\ }\href@noop {} {\bibfield  {journal} {\bibinfo  {journal} {Journal of the Science of Food and Agriculture}\ }\textbf {\bibinfo {volume} {100}},\ \bibinfo {pages} {2807} (\bibinfo {year} {2020})}\BibitemShut {NoStop}%
\bibitem [{\citenamefont {Bormashenko}(2011)}]{bormashenko2011liquid}%
  \BibitemOpen
  \bibfield  {author} {\bibinfo {author} {\bibfnamefont {E.}~\bibnamefont {Bormashenko}},\ }\bibfield  {title} {\bibinfo {title} {Liquid marbles: properties and applications},\ }\href@noop {} {\bibfield  {journal} {\bibinfo  {journal} {Current Opinion in Colloid \& Interface Science}\ }\textbf {\bibinfo {volume} {16}},\ \bibinfo {pages} {266} (\bibinfo {year} {2011})}\BibitemShut {NoStop}%
\bibitem [{\citenamefont {Subramaniam}\ \emph {et~al.}(2006)\citenamefont {Subramaniam}, \citenamefont {Abkarian}, \citenamefont {Mahadevan},\ and\ \citenamefont {Stone}}]{subramaniam2006mechanics}%
  \BibitemOpen
  \bibfield  {author} {\bibinfo {author} {\bibfnamefont {A.~B.}\ \bibnamefont {Subramaniam}}, \bibinfo {author} {\bibfnamefont {M.}~\bibnamefont {Abkarian}}, \bibinfo {author} {\bibfnamefont {L.}~\bibnamefont {Mahadevan}},\ and\ \bibinfo {author} {\bibfnamefont {H.~A.}\ \bibnamefont {Stone}},\ }\bibfield  {title} {\bibinfo {title} {Mechanics of interfacial composite materials},\ }\href@noop {} {\bibfield  {journal} {\bibinfo  {journal} {Langmuir}\ }\textbf {\bibinfo {volume} {22}},\ \bibinfo {pages} {10204} (\bibinfo {year} {2006})}\BibitemShut {NoStop}%
\bibitem [{\citenamefont {Dan}(2012)}]{dan2012transport}%
  \BibitemOpen
  \bibfield  {author} {\bibinfo {author} {\bibfnamefont {N.}~\bibnamefont {Dan}},\ }\bibfield  {title} {\bibinfo {title} {Transport through self-assembled colloidal shells (colloidosomes)},\ }\href@noop {} {\bibfield  {journal} {\bibinfo  {journal} {Current opinion in colloid \& interface science}\ }\textbf {\bibinfo {volume} {17}},\ \bibinfo {pages} {141} (\bibinfo {year} {2012})}\BibitemShut {NoStop}%
\bibitem [{\citenamefont {Sacca}\ \emph {et~al.}(2008)\citenamefont {Sacca}, \citenamefont {Drelich}, \citenamefont {Gomez}, \citenamefont {Pezron},\ and\ \citenamefont {Clausse}}]{sacca2008composition}%
  \BibitemOpen
  \bibfield  {author} {\bibinfo {author} {\bibfnamefont {L.}~\bibnamefont {Sacca}}, \bibinfo {author} {\bibfnamefont {A.}~\bibnamefont {Drelich}}, \bibinfo {author} {\bibfnamefont {F.}~\bibnamefont {Gomez}}, \bibinfo {author} {\bibfnamefont {I.}~\bibnamefont {Pezron}},\ and\ \bibinfo {author} {\bibfnamefont {D.}~\bibnamefont {Clausse}},\ }\bibfield  {title} {\bibinfo {title} {Composition ripening in mixed water-in-oil emulsions stabilized with solid particles},\ }\href@noop {} {\bibfield  {journal} {\bibinfo  {journal} {Journal of dispersion science and technology}\ }\textbf {\bibinfo {volume} {29}},\ \bibinfo {pages} {948} (\bibinfo {year} {2008})}\BibitemShut {NoStop}%
\bibitem [{\citenamefont {Yong}\ \emph {et~al.}(2016)\citenamefont {Yong}, \citenamefont {Qin},\ and\ \citenamefont {Singler}}]{yong2016nanoparticle}%
  \BibitemOpen
  \bibfield  {author} {\bibinfo {author} {\bibfnamefont {X.}~\bibnamefont {Yong}}, \bibinfo {author} {\bibfnamefont {S.}~\bibnamefont {Qin}},\ and\ \bibinfo {author} {\bibfnamefont {T.~J.}\ \bibnamefont {Singler}},\ }\bibfield  {title} {\bibinfo {title} {Nanoparticle-mediated evaporation at liquid--vapor interfaces},\ }\href@noop {} {\bibfield  {journal} {\bibinfo  {journal} {Extreme Mechanics Letters}\ }\textbf {\bibinfo {volume} {7}},\ \bibinfo {pages} {90} (\bibinfo {year} {2016})}\BibitemShut {NoStop}%
\bibitem [{\citenamefont {Thompson}\ \emph {et~al.}(2010)\citenamefont {Thompson}, \citenamefont {Armes}, \citenamefont {Howse}, \citenamefont {Ebbens}, \citenamefont {Ahmad}, \citenamefont {Zaidi}, \citenamefont {York},\ and\ \citenamefont {Burdis}}]{thompson2010covalently}%
  \BibitemOpen
  \bibfield  {author} {\bibinfo {author} {\bibfnamefont {K.~L.}\ \bibnamefont {Thompson}}, \bibinfo {author} {\bibfnamefont {S.}~\bibnamefont {Armes}}, \bibinfo {author} {\bibfnamefont {J.}~\bibnamefont {Howse}}, \bibinfo {author} {\bibfnamefont {S.}~\bibnamefont {Ebbens}}, \bibinfo {author} {\bibfnamefont {I.}~\bibnamefont {Ahmad}}, \bibinfo {author} {\bibfnamefont {J.}~\bibnamefont {Zaidi}}, \bibinfo {author} {\bibfnamefont {D.}~\bibnamefont {York}},\ and\ \bibinfo {author} {\bibfnamefont {J.}~\bibnamefont {Burdis}},\ }\bibfield  {title} {\bibinfo {title} {Covalently cross-linked colloidosomes},\ }\href@noop {} {\bibfield  {journal} {\bibinfo  {journal} {Macromolecules}\ }\textbf {\bibinfo {volume} {43}},\ \bibinfo {pages} {10466} (\bibinfo {year} {2010})}\BibitemShut {NoStop}%
\bibitem [{\citenamefont {Schr{\"o}der}\ \emph {et~al.}(2019)\citenamefont {Schr{\"o}der}, \citenamefont {Sprakel}, \citenamefont {Boerkamp}, \citenamefont {Schro{\"e}n},\ and\ \citenamefont {Berton-Carabin}}]{schroder2019can}%
  \BibitemOpen
  \bibfield  {author} {\bibinfo {author} {\bibfnamefont {A.}~\bibnamefont {Schr{\"o}der}}, \bibinfo {author} {\bibfnamefont {J.}~\bibnamefont {Sprakel}}, \bibinfo {author} {\bibfnamefont {W.}~\bibnamefont {Boerkamp}}, \bibinfo {author} {\bibfnamefont {K.}~\bibnamefont {Schro{\"e}n}},\ and\ \bibinfo {author} {\bibfnamefont {C.~C.}\ \bibnamefont {Berton-Carabin}},\ }\bibfield  {title} {\bibinfo {title} {Can we prevent lipid oxidation in emulsions by using fat-based pickering particles?},\ }\href@noop {} {\bibfield  {journal} {\bibinfo  {journal} {Food Research International}\ }\textbf {\bibinfo {volume} {120}},\ \bibinfo {pages} {352} (\bibinfo {year} {2019})}\BibitemShut {NoStop}%
\bibitem [{\citenamefont {Prakash}\ \emph {et~al.}(2025)\citenamefont {Prakash}, \citenamefont {Krolis}, \citenamefont {Marin},\ and\ \citenamefont {Botto}}]{prakash2025evaporation}%
  \BibitemOpen
  \bibfield  {author} {\bibinfo {author} {\bibfnamefont {S.}~\bibnamefont {Prakash}}, \bibinfo {author} {\bibfnamefont {E.}~\bibnamefont {Krolis}}, \bibinfo {author} {\bibfnamefont {A.}~\bibnamefont {Marin}},\ and\ \bibinfo {author} {\bibfnamefont {L.}~\bibnamefont {Botto}},\ }\bibfield  {title} {\bibinfo {title} {Evaporation driven buckling of a drop laden with graphene oxide nanosheets},\ }\href@noop {} {\bibfield  {journal} {\bibinfo  {journal} {Soft Matter}\ } (\bibinfo {year} {2025})}\BibitemShut {NoStop}%
\bibitem [{\citenamefont {Yow}\ and\ \citenamefont {Routh}(2009)}]{yow2009release}%
  \BibitemOpen
  \bibfield  {author} {\bibinfo {author} {\bibfnamefont {H.~N.}\ \bibnamefont {Yow}}\ and\ \bibinfo {author} {\bibfnamefont {A.~F.}\ \bibnamefont {Routh}},\ }\bibfield  {title} {\bibinfo {title} {Release profiles of encapsulated actives from colloidosomes sintered for various durations},\ }\href@noop {} {\bibfield  {journal} {\bibinfo  {journal} {Langmuir}\ }\textbf {\bibinfo {volume} {25}},\ \bibinfo {pages} {159} (\bibinfo {year} {2009})}\BibitemShut {NoStop}%
\bibitem [{\citenamefont {Liu}\ \emph {et~al.}(2024)\citenamefont {Liu}, \citenamefont {Xu}, \citenamefont {Portela},\ and\ \citenamefont {Garbin}}]{liu2024diffusion}%
  \BibitemOpen
  \bibfield  {author} {\bibinfo {author} {\bibfnamefont {Y.}~\bibnamefont {Liu}}, \bibinfo {author} {\bibfnamefont {M.}~\bibnamefont {Xu}}, \bibinfo {author} {\bibfnamefont {L.~M.}\ \bibnamefont {Portela}},\ and\ \bibinfo {author} {\bibfnamefont {V.}~\bibnamefont {Garbin}},\ }\bibfield  {title} {\bibinfo {title} {Diffusion across particle-laden interfaces in pickering droplets},\ }\href@noop {} {\bibfield  {journal} {\bibinfo  {journal} {Soft Matter}\ }\textbf {\bibinfo {volume} {20}},\ \bibinfo {pages} {94} (\bibinfo {year} {2024})}\BibitemShut {NoStop}%
\bibitem [{\citenamefont {Suzuki}\ and\ \citenamefont {Maeda}(1968)}]{suzuki1968mechanism}%
  \BibitemOpen
  \bibfield  {author} {\bibinfo {author} {\bibfnamefont {M.}~\bibnamefont {Suzuki}}\ and\ \bibinfo {author} {\bibfnamefont {S.}~\bibnamefont {Maeda}},\ }\bibfield  {title} {\bibinfo {title} {On the mechanism of drying of granular beds mass transfer from discontinuous source},\ }\href@noop {} {\bibfield  {journal} {\bibinfo  {journal} {Journal of chemical engineering of Japan}\ }\textbf {\bibinfo {volume} {1}},\ \bibinfo {pages} {26} (\bibinfo {year} {1968})}\BibitemShut {NoStop}%
\bibitem [{\citenamefont {Shahraeeni}\ \emph {et~al.}(2012)\citenamefont {Shahraeeni}, \citenamefont {Lehmann},\ and\ \citenamefont {Or}}]{shahraeeni2012coupling}%
  \BibitemOpen
  \bibfield  {author} {\bibinfo {author} {\bibfnamefont {E.}~\bibnamefont {Shahraeeni}}, \bibinfo {author} {\bibfnamefont {P.}~\bibnamefont {Lehmann}},\ and\ \bibinfo {author} {\bibfnamefont {D.}~\bibnamefont {Or}},\ }\bibfield  {title} {\bibinfo {title} {Coupling of evaporative fluxes from drying porous surfaces with air boundary layer: Characteristics of evaporation from discrete pores},\ }\href@noop {} {\bibfield  {journal} {\bibinfo  {journal} {Water Resources Research}\ }\textbf {\bibinfo {volume} {48}} (\bibinfo {year} {2012})}\BibitemShut {NoStop}%
\bibitem [{\citenamefont {Berg}(1993)}]{berg1993random}%
  \BibitemOpen
  \bibfield  {author} {\bibinfo {author} {\bibfnamefont {H.~C.}\ \bibnamefont {Berg}},\ }\href@noop {} {\emph {\bibinfo {title} {Random walks in biology}}}\ (\bibinfo  {publisher} {Princeton University Press},\ \bibinfo {year} {1993})\BibitemShut {NoStop}%
\bibitem [{\citenamefont {Zwanzig}(1992)}]{zwanzig1992diffusion}%
  \BibitemOpen
  \bibfield  {author} {\bibinfo {author} {\bibfnamefont {R.}~\bibnamefont {Zwanzig}},\ }\bibfield  {title} {\bibinfo {title} {Diffusion past an entropy barrier},\ }\href@noop {} {\bibfield  {journal} {\bibinfo  {journal} {The Journal of Physical Chemistry}\ }\textbf {\bibinfo {volume} {96}},\ \bibinfo {pages} {3926} (\bibinfo {year} {1992})}\BibitemShut {NoStop}%
\bibitem [{\citenamefont {Reguera}\ and\ \citenamefont {Rubi}(2001)}]{reguera2001kinetic}%
  \BibitemOpen
  \bibfield  {author} {\bibinfo {author} {\bibfnamefont {D.}~\bibnamefont {Reguera}}\ and\ \bibinfo {author} {\bibfnamefont {J.}~\bibnamefont {Rubi}},\ }\bibfield  {title} {\bibinfo {title} {Kinetic equations for diffusion in the presence of entropic barriers},\ }\href@noop {} {\bibfield  {journal} {\bibinfo  {journal} {Physical Review E}\ }\textbf {\bibinfo {volume} {64}},\ \bibinfo {pages} {061106} (\bibinfo {year} {2001})}\BibitemShut {NoStop}%
\bibitem [{\citenamefont {Kalinay}\ and\ \citenamefont {Percus}(2006)}]{kalinay2006corrections}%
  \BibitemOpen
  \bibfield  {author} {\bibinfo {author} {\bibfnamefont {P.}~\bibnamefont {Kalinay}}\ and\ \bibinfo {author} {\bibfnamefont {J.}~\bibnamefont {Percus}},\ }\bibfield  {title} {\bibinfo {title} {Corrections to the fick-jacobs equation},\ }\href@noop {} {\bibfield  {journal} {\bibinfo  {journal} {Physical Review E—Statistical, Nonlinear, and Soft Matter Physics}\ }\textbf {\bibinfo {volume} {74}},\ \bibinfo {pages} {041203} (\bibinfo {year} {2006})}\BibitemShut {NoStop}%
\bibitem [{\citenamefont {Jacobs}(1935)}]{jacobs1935diffusion}%
  \BibitemOpen
  \bibfield  {author} {\bibinfo {author} {\bibfnamefont {M.~H.}\ \bibnamefont {Jacobs}},\ }\href@noop {} {\emph {\bibinfo {title} {Diffusion processes}}}\ (\bibinfo  {publisher} {Springer},\ \bibinfo {year} {1935})\BibitemShut {NoStop}%
\bibitem [{\citenamefont {Carslaw}\ and\ \citenamefont {Jaeger}(1959)}]{CarslawJaeger1959}%
  \BibitemOpen
  \bibfield  {author} {\bibinfo {author} {\bibfnamefont {H.~S.}\ \bibnamefont {Carslaw}}\ and\ \bibinfo {author} {\bibfnamefont {J.~C.}\ \bibnamefont {Jaeger}},\ }\href@noop {} {\emph {\bibinfo {title} {Conduction of Heat in Solids}}},\ \bibinfo {edition} {2nd}\ ed.\ (\bibinfo  {publisher} {Oxford University Press},\ \bibinfo {address} {Oxford},\ \bibinfo {year} {1959})\BibitemShut {NoStop}%
\bibitem [{\citenamefont {Laborie}\ \emph {et~al.}(2013)\citenamefont {Laborie}, \citenamefont {Lachauss{\'e}e}, \citenamefont {Lorenceau},\ and\ \citenamefont {Rouyer}}]{laborie2013coatings}%
  \BibitemOpen
  \bibfield  {author} {\bibinfo {author} {\bibfnamefont {B.}~\bibnamefont {Laborie}}, \bibinfo {author} {\bibfnamefont {F.}~\bibnamefont {Lachauss{\'e}e}}, \bibinfo {author} {\bibfnamefont {E.}~\bibnamefont {Lorenceau}},\ and\ \bibinfo {author} {\bibfnamefont {F.}~\bibnamefont {Rouyer}},\ }\bibfield  {title} {\bibinfo {title} {How coatings with hydrophobic particles may change the drying of water droplets: incompressible surface versus porous media effects},\ }\href@noop {} {\bibfield  {journal} {\bibinfo  {journal} {Soft Matter}\ }\textbf {\bibinfo {volume} {9}},\ \bibinfo {pages} {4822} (\bibinfo {year} {2013})}\BibitemShut {NoStop}%
\bibitem [{\citenamefont {Asaumi}\ \emph {et~al.}(2020)\citenamefont {Asaumi}, \citenamefont {Rey}, \citenamefont {Oyama}, \citenamefont {Vogel}, \citenamefont {Hirai}, \citenamefont {Nakamura},\ and\ \citenamefont {Fujii}}]{asaumi2020effect}%
  \BibitemOpen
  \bibfield  {author} {\bibinfo {author} {\bibfnamefont {Y.}~\bibnamefont {Asaumi}}, \bibinfo {author} {\bibfnamefont {M.}~\bibnamefont {Rey}}, \bibinfo {author} {\bibfnamefont {K.}~\bibnamefont {Oyama}}, \bibinfo {author} {\bibfnamefont {N.}~\bibnamefont {Vogel}}, \bibinfo {author} {\bibfnamefont {T.}~\bibnamefont {Hirai}}, \bibinfo {author} {\bibfnamefont {Y.}~\bibnamefont {Nakamura}},\ and\ \bibinfo {author} {\bibfnamefont {S.}~\bibnamefont {Fujii}},\ }\bibfield  {title} {\bibinfo {title} {Effect of stabilizing particle size on the structure and properties of liquid marbles},\ }\href@noop {} {\bibfield  {journal} {\bibinfo  {journal} {Langmuir}\ }\textbf {\bibinfo {volume} {36}},\ \bibinfo {pages} {13274} (\bibinfo {year} {2020})}\BibitemShut {NoStop}%
\bibitem [{\citenamefont {Saczek}\ \emph {et~al.}(2024)\citenamefont {Saczek}, \citenamefont {Murphy}, \citenamefont {Zivkovic}, \citenamefont {Putranto},\ and\ \citenamefont {Pramana}}]{saczek2024impact}%
  \BibitemOpen
  \bibfield  {author} {\bibinfo {author} {\bibfnamefont {J.}~\bibnamefont {Saczek}}, \bibinfo {author} {\bibfnamefont {K.}~\bibnamefont {Murphy}}, \bibinfo {author} {\bibfnamefont {V.}~\bibnamefont {Zivkovic}}, \bibinfo {author} {\bibfnamefont {A.}~\bibnamefont {Putranto}},\ and\ \bibinfo {author} {\bibfnamefont {S.~S.}\ \bibnamefont {Pramana}},\ }\bibfield  {title} {\bibinfo {title} {Impact of coating particles on liquid marble lifetime: reactor engineering approach to evaporation},\ }\href@noop {} {\bibfield  {journal} {\bibinfo  {journal} {Soft Matter}\ }\textbf {\bibinfo {volume} {20}},\ \bibinfo {pages} {5822} (\bibinfo {year} {2024})}\BibitemShut {NoStop}%
\bibitem [{\citenamefont {Kim}\ \emph {et~al.}(2007)\citenamefont {Kim}, \citenamefont {Fern{\'a}ndez-Nieves}, \citenamefont {Dan}, \citenamefont {Utada}, \citenamefont {Marquez},\ and\ \citenamefont {Weitz}}]{kim2007colloidal}%
  \BibitemOpen
  \bibfield  {author} {\bibinfo {author} {\bibfnamefont {J.-W.}\ \bibnamefont {Kim}}, \bibinfo {author} {\bibfnamefont {A.}~\bibnamefont {Fern{\'a}ndez-Nieves}}, \bibinfo {author} {\bibfnamefont {N.}~\bibnamefont {Dan}}, \bibinfo {author} {\bibfnamefont {A.~S.}\ \bibnamefont {Utada}}, \bibinfo {author} {\bibfnamefont {M.}~\bibnamefont {Marquez}},\ and\ \bibinfo {author} {\bibfnamefont {D.~A.}\ \bibnamefont {Weitz}},\ }\bibfield  {title} {\bibinfo {title} {Colloidal assembly route for responsive colloidosomes with tunable permeability},\ }\href@noop {} {\bibfield  {journal} {\bibinfo  {journal} {Nano Letters}\ }\textbf {\bibinfo {volume} {7}},\ \bibinfo {pages} {2876} (\bibinfo {year} {2007})}\BibitemShut {NoStop}%
\bibitem [{\citenamefont {Sj{\"o}{\"o}}\ \emph {et~al.}(2015)\citenamefont {Sj{\"o}{\"o}}, \citenamefont {Emek}, \citenamefont {Hall}, \citenamefont {Rayner},\ and\ \citenamefont {Wahlgren}}]{sjoo2015barrier}%
  \BibitemOpen
  \bibfield  {author} {\bibinfo {author} {\bibfnamefont {M.}~\bibnamefont {Sj{\"o}{\"o}}}, \bibinfo {author} {\bibfnamefont {S.~C.}\ \bibnamefont {Emek}}, \bibinfo {author} {\bibfnamefont {T.}~\bibnamefont {Hall}}, \bibinfo {author} {\bibfnamefont {M.}~\bibnamefont {Rayner}},\ and\ \bibinfo {author} {\bibfnamefont {M.}~\bibnamefont {Wahlgren}},\ }\bibfield  {title} {\bibinfo {title} {Barrier properties of heat treated starch pickering emulsions},\ }\href@noop {} {\bibfield  {journal} {\bibinfo  {journal} {Journal of colloid and interface science}\ }\textbf {\bibinfo {volume} {450}},\ \bibinfo {pages} {182} (\bibinfo {year} {2015})}\BibitemShut {NoStop}%
\bibitem [{\citenamefont {Jia}\ \emph {et~al.}(2023)\citenamefont {Jia}, \citenamefont {Liu}, \citenamefont {Sun},\ and\ \citenamefont {Wang}}]{jia2023efficient}%
  \BibitemOpen
  \bibfield  {author} {\bibinfo {author} {\bibfnamefont {J.}~\bibnamefont {Jia}}, \bibinfo {author} {\bibfnamefont {R.-K.}\ \bibnamefont {Liu}}, \bibinfo {author} {\bibfnamefont {Q.}~\bibnamefont {Sun}},\ and\ \bibinfo {author} {\bibfnamefont {J.-X.}\ \bibnamefont {Wang}},\ }\bibfield  {title} {\bibinfo {title} {Efficient construction of ph-stimuli-responsive colloidosomes with high encapsulation efficiency},\ }\href@noop {} {\bibfield  {journal} {\bibinfo  {journal} {Langmuir}\ }\textbf {\bibinfo {volume} {39}},\ \bibinfo {pages} {17808} (\bibinfo {year} {2023})}\BibitemShut {NoStop}%
\bibitem [{\citenamefont {Abkarian}\ \emph {et~al.}(2007)\citenamefont {Abkarian}, \citenamefont {Subramaniam}, \citenamefont {Kim}, \citenamefont {Larsen}, \citenamefont {Yang},\ and\ \citenamefont {Stone}}]{abkarian2007dissolution}%
  \BibitemOpen
  \bibfield  {author} {\bibinfo {author} {\bibfnamefont {M.}~\bibnamefont {Abkarian}}, \bibinfo {author} {\bibfnamefont {A.~B.}\ \bibnamefont {Subramaniam}}, \bibinfo {author} {\bibfnamefont {S.-H.}\ \bibnamefont {Kim}}, \bibinfo {author} {\bibfnamefont {R.~J.}\ \bibnamefont {Larsen}}, \bibinfo {author} {\bibfnamefont {.~f. S.-M.}\ \bibnamefont {Yang}},\ and\ \bibinfo {author} {\bibfnamefont {H.~A.}\ \bibnamefont {Stone}},\ }\bibfield  {title} {\bibinfo {title} {Dissolution arrest and stability of particle-covered bubbles},\ }\href@noop {} {\bibfield  {journal} {\bibinfo  {journal} {Physical review letters}\ }\textbf {\bibinfo {volume} {99}},\ \bibinfo {pages} {188301} (\bibinfo {year} {2007})}\BibitemShut {NoStop}%
\bibitem [{\citenamefont {Taccoen}\ \emph {et~al.}(2016)\citenamefont {Taccoen}, \citenamefont {Lequeux}, \citenamefont {Gunes},\ and\ \citenamefont {Baroud}}]{taccoen2016probing}%
  \BibitemOpen
  \bibfield  {author} {\bibinfo {author} {\bibfnamefont {N.}~\bibnamefont {Taccoen}}, \bibinfo {author} {\bibfnamefont {F.}~\bibnamefont {Lequeux}}, \bibinfo {author} {\bibfnamefont {D.~Z.}\ \bibnamefont {Gunes}},\ and\ \bibinfo {author} {\bibfnamefont {C.~N.}\ \bibnamefont {Baroud}},\ }\bibfield  {title} {\bibinfo {title} {Probing the mechanical strength of an armored bubble and its implication to particle-stabilized foams},\ }\href@noop {} {\bibfield  {journal} {\bibinfo  {journal} {Physical Review X}\ }\textbf {\bibinfo {volume} {6}},\ \bibinfo {pages} {011010} (\bibinfo {year} {2016})}\BibitemShut {NoStop}%
\bibitem [{\citenamefont {Poulichet}\ and\ \citenamefont {Garbin}(2015)}]{poulichet2015cooling}%
  \BibitemOpen
  \bibfield  {author} {\bibinfo {author} {\bibfnamefont {V.}~\bibnamefont {Poulichet}}\ and\ \bibinfo {author} {\bibfnamefont {V.}~\bibnamefont {Garbin}},\ }\bibfield  {title} {\bibinfo {title} {Cooling particle-coated bubbles: Destabilization beyond dissolution arrest},\ }\href@noop {} {\bibfield  {journal} {\bibinfo  {journal} {Langmuir}\ }\textbf {\bibinfo {volume} {31}},\ \bibinfo {pages} {12035} (\bibinfo {year} {2015})}\BibitemShut {NoStop}%
\bibitem [{\citenamefont {van Overveld}\ and\ \citenamefont {Garbin}(2025)}]{vanoverveld2025dataset}%
  \BibitemOpen
  \bibfield  {author} {\bibinfo {author} {\bibfnamefont {T.~J. J.~M.}\ \bibnamefont {van Overveld}}\ and\ \bibinfo {author} {\bibfnamefont {V.}~\bibnamefont {Garbin}},\ }\href {https://doi.org/10.4121/c147775d-af0e-4c8e-926a-ffd1ee1a8e0c} {\bibinfo {title} {Numerical method and dataset underlying the letter: Scaling regimes for unsteady diffusion across particle-stabilized fluid interfaces}} (\bibinfo {year} {2025})\BibitemShut {NoStop}%
\end{thebibliography}%

%

\end{document}